\documentclass[11pt]{article}

\usepackage{xy}
\xyoption{all}

\makeatletter

\def\qed{\ifvmode\removelastskip\fi
{\unskip\nobreak\hfil\penalty50\hbox{}\nobreak\hfil
\hbox{\vrule height1.2ex width1.2ex}\parfillskip=0pt
\finalhyphendemerits=0 \par\smallskip}}

\def\dif{{\rm d}}
\def\deriv{\@ifnextchar[{\@deriv}{\@deriv[]}}
   \def\@deriv[#1]#2#3{\mathchoice%
{{\dif^{#1}#2\over\dif{#3}^{#1}}}{{\dif^{#1}#2/\dif{#3}^{#1}}}%
{{\dif^{#1}#2\over\dif{#3}^{#1}}}{{\dif^{#1}#2/\dif{#3}^{#1}}}}
\def\derpar#1#2{\mathchoice%
{{\partial#1\over\partial#2}}{{\partial#1/\partial#2}}%
{{\partial#1\over\partial#2}}{{\partial#1/\partial#2}}}

\def\restric#1#2{{\left. #1 \right|_{#2}}}

\def\secteqno{\@addtoreset{equation}{section}%
\def\theequation{\thesection.\arabic{equation}}}

\newcounter{subequation}
\def\thesubequation{\alph{subequation}}
\def\sneqnarray{\stepcounter{equation}\let\@currentlabel=\theequation
\setcounter{subequation}{1}
\def\@eqnnum{{\rm (\theequation.\thesubequation)}}
\global\@eqcnt\z@\tabskip\@centering\let\\=\@eqncr\let\@@eqncr=\@@sneqncr
$$\halign to \displaywidth\bgroup\@eqnsel\hskip\@centering
 $\displaystyle\tabskip\z@{##}$&\global\@eqcnt\@ne
 \hskip 2\arraycolsep \hfil${##}$\hfil
 &\global\@eqcnt\tw@ \hskip 2\arraycolsep $\displaystyle\tabskip\z@{##}$\hfil
  \tabskip\@centering&\llap{##}\tabskip\z@\cr}
\def\endsneqnarray{\@@sneqncr\egroup $$\global\@ignoretrue}
\def\@@sneqncr{\let\@tempa\relax
   \ifcase\@eqcnt \def\@tempa{& & &}\or \def\@tempa{& &}
   \else \def\@tempa{&}\fi
     \@tempa \if@eqnsw\@eqnnum\stepcounter{subequation}\fi
     \global\@eqnswtrue\global\@eqcnt\z@\cr}

\def\nobiblabels{\def\@lbibitem[##1]##2{\@bibitem{##2}}}
\makeatother

\def\ben{\begin{enumerate}}
\def\een{\end{enumerate}}
\def\beq{\begin{equation}}
\def\eeq{\end{equation}}
\def\bea{\begin{eqnarray}}
\def\eea{\end{eqnarray}}
\def\beann{\begin{eqnarray*}}
\def\eeann{\end{eqnarray*}}
\def\beasn{\begin{sneqnarray}}
\def\eeasn{\end{sneqnarray}}

\newtheorem{prop}{Proposition}

\let\ds=\displaystyle
\def\buildord#1\over#2{\mathord{\mathop{\kern0pt #2}\limits^{#1}}}

\def\Real{{\bf R}}

\def\Ker{\mathop{\rm Ker}\nolimits}

\def\transp#1{{}^{t}\kern-.15em\relax#1}
\def\Tens^#1{\buildord #1\over\otimes }

\def\Img{\mathop{\rm Im}\nolimits}

\def\Aff{\mathcal{A}}
\def\Lin{\mathcal{L}}
\def\R{\mathbf{R}}
\def\hh{{\hat{\,}}}

\def\Cinfty{{\rm C}^\infty}

\def\feble#1{\mathrel{\mathop{\simeq}\limits_{#1}}}

\def\lop{\!\cdot\!}

\def\Tan{{\rm T}}
\def\Ver{{\rm V}}
\def\Jet{{\rm J}}
\def\jet{{\rm j}}

\def\tanvec#1{{\partial \over \partial #1}}

\font\gotic=eufm10 scaled \magstephalf
\def\Vect{\hbox{\gotic X}}

\def\vend{S}

\def\calD{\mathcal{D}}

\def\proof{\noindent{\itshape Proof}.\quad}

\def\tabaddress#1{{\small\sffamily\slshape\begin{tabular}[t]{c}#1
\\[1.2ex]\end{tabular}}}
\def\UPCMAT{Departament de Matem\`atica Aplicada IV\\
   Universitat Polit\`ecnica de Catalunya\\
   Campus Nord UPC, edifici C3\\
   C/ Jordi Girona~1,
   08034 Barcelona\\
   Catalonia, Spain}

\title{\sffamily Time-dependent singular differential equations}

\author{\sffamily
Xavier Gr\`acia and Rub\'en Mart\'{\i}n
\\[1mm]
\tabaddress{\UPCMAT}
\\[2mm]
\small\sffamily\slshape emails: xgracia@mat.upc.es, rubenmg@mat.upc.es
}

\date{\sffamily 25 October 2004}

\begin{document}

\maketitle
\thispagestyle{empty}

\begin{abstract}
\parindent0pt\noindent
A geometric framework for describing and solving 
time-dependent implicit differential equations 
$F(t,x,x')=0$ is studied, 
paying special attention to the linearly singular case, 
where $F$ is affine in the velocities: 
$A(t,x)x' = b(t,x)$. 
This framework is based on the jet bundle of 
a time-dependent configuration space, 
and is an extension of the geometric framework 
of the autonomous case. 
When $A$ is a singular matrix, 
the solutions can be obtained by means of constraint algorithms,
either directly or 
through an equivalent autonomous system 
that can be constructed using the vector hull functor of affine spaces. 
As applications, we consider the jet bundle description of
time-dependent lagrangian systems and the
Skinner--Rusk formulation of time-dependent mechanics.



\end{abstract}

\clearpage
\section{Introduction}

It is known that the dynamics of an
autonomous mechanical system given
by a lagrangian function 
$L\colon\Tan Q\to\Real$
is described by equations of motion of the type
$W(q,\dot q)\,\ddot q = \alpha(q,\dot q)$;
writing them as a first-order differential equation 
on $\Tan Q$, we obtain
$$
A(x)\dot x = b(x),
$$
where the matrix~$A$ is singular 
if the lagrangian is not regular
(see for instance \cite{Car-sing}).
Indeed, such differential equations appear in many other applications 
as control theory, circuit theory, engineering, \ldots 
---see some references in
\cite{GMR-diffeq}.
 
A general framework to deal with differential
equations of this type was given in 
\cite{GP-unif}
and developed in 
\cite{GP-gener,GP-symmsing}:
the so-called {\itshape linearly singular systems}.
Since the matrix~$A$ may be singular,
the system may not have solutions passing through 
each point of~$\Tan Q$,
and the solutions may not be unique.
To solve this problem,
a constraint algorithm has to be performed.
If the algorithm ends,
one obtains a submanifold of~$\Tan Q$
where there exist solutions of the system.
In fact, this algorithm is a generalization
of the presymplectic constraint algorithm 
\cite{GNH-pres}, 
which is in its turn a geometric version 
of the Dirac--Bergmann theory of constrained systems.

\medskip

The main purpose of this paper is
to extend these results to the time-dependent case.
We study the geometric framework of
time-dependent first-order implicit differential equations, 
$$
F(t,x,\dot x) = 0 ,
$$
and the linearly singular case, 
when $F$ is affine in the velocities, 
$$
A(t,x)\dot x = b(t,x),
$$
where~$A$ is a possibly singular matrix.

Our model for the time-dependent configuration space, 
rather than a trivial product $M = \Real \times Q$, 
is a fibre bundle  
\cite{Sau-jets}
$\rho \colon M \to \Real$, 
where the base $\Real$ contains the time variable. 
Such an $M$ is isomorphic to a product $\Real \times Q$, 
but in practical applications 
there may not be a privileged trivialization, 
and a possible extension to deal with field theory 
of course should not be based on a trivial bundle. 
Some references about time-dependent lagrangian systems are
\cite{EMR-time,CF-time,CLM-algorithm} 
in the product case and 
\cite{Kru-LNM,MPV-Legendre,MS-mech,LMMMR-sing}
in the fibre bundle case; 
see also references therein. 
Time-dependent systems in general are studied in many books, 
as for instance 
\cite{AM-mech,Olv-diffeq}.

The main difference between 
the autonomous and the non-autonomous case 
is the usage of tangent bundles and jet bundles 
respectively. 
To describe an autonomous differential equation on
the configuration space~$Q$ 
we use the tangent bundle $\Tan Q$, 
which is a vector bundle.
To describe a non-autonomous differential equation on
the time-dependent configuration space~$M$, 
we use its jet bundle $\Jet^1 \rho$, 
which is an affine bundle over~$M$.
To describe a linearly singular equation on~$M$ 
we will use affine morphisms defined on this affine bundle.

We will also propose a constraint algorithm
for time-dependent singular systems.
This algorithm is the natural generalization
of the algorithm for the autonomous case 
---both in the general implicit
\cite{RR-94,MMT-95} 
and linearly singular cases
\cite{GP-unif, GP-gener}. 
The case of an implicit equation in a product $M = \Real \times Q$
has already been discussed in
\cite{Del-tesi}.
It is worth noting that 
constraint algorithms for some time-dependent systems 
have been described in several papers, 
as for instance 
\cite{CF-time,CLM-algorithm,ILMM-Dir,LMM-constraint,LMMMR-sing,Vig}. 
All these systems are of linearly singular type, 
so they are included within our framework. 

When studying a time-dependent differential equation, 
sometimes it is useful
to convert it into an equivalent time-independent one. 
This is even more interesting for implicit equations;  
for instance, the constraint algorithm 
for the autonomous case 
is easier to implement than
for the non-autonomous case, 
because of the fact that vector fields
instead of jet fields are used
to obtain the constraint functions.

Therefore, 
we will examine the possibility of associating an autonomous 
linearly singular system with a time-dependent one,
so that the solutions 
of both systems will be in correspondence. 
Essentially, we use the canonical inclusion 
of $\Jet^1 \rho$ into $\Tan M$. 
In order to perform this association,
we propose two different strategies.
One possibility is to choose a connection
on the jet bundle to induce a splitting
of the tangent bundle.
The other possibility, which does not make use
of any choice, is based on the notion of
vector hull of an affine space or an affine bundle. 
The main idea is that any affine space $A$ 
can be canonically embedded as a hyperplane 
in a vector space~$\hat A$ ---the vector hull of~$A$; 
with this immersion affine maps can be homogenized, 
that is, converted into linear maps. 

As we have pointed out at the beginning, 
our main motivation comes from 
Euler--Lagrange equations and mechanical systems, 
where equations of motion are of second order. 
So it is interesting to extend the preceding study 
to second-order implicit and linearly singular equations:
$$
F(t,x,\dot x,\ddot x) = 0 , 
\qquad
A(t,x,\dot x)\ddot x = b(t,x,\dot x) .
$$

As applications of the formalism, 
we give two descriptions of 
time-dependent mechanical systems, 
in the form of 
time-dependent singular lagrangian systems
and in the mixed velocity-momentum description 
(sometimes called Skinner--Rusk formulation
\cite{Ski-mixt,SR-mixt}) 
of time-dependent mechanics 
\cite{CMC-SR}.
As a concrete example, 
we also study a pendulum of variable length.

\medskip

The paper is organized as follows.
In section~2 we study the geometric formulation of 
time-dependent differential equations, 
either in the implicit and in the linearly singular case. 
In section~3 we describe constraint algorithms for both cases. 
In section~4 we present two constructions of an autonomous system 
associated with a given time-dependent system. 
The extension to second-order equations is presented in section~5. 
Applications to singular lagrangian mechanics are presented 
in section~6, 
and section~7 is devoted to an example. 
Finally, there is an appendix about vector hulls of affine spaces 
and affine bundles.

\section{Time-dependent systems}

In this section we discuss first-order time-dependent
singular differential equations.
As a model of time-dependent configuration space,
we take a fibre bundle
$\rho\colon M\to \Real$
over the real line
(though more general settings could be also considered).

The appropriate geometric framework to deal with derivatives
is that of jet bundles,
so we will begin by giving some facts and notation about them
---see for instance
\cite{Sau-jets}.

\subsection{First-order jet bundles}

We denote by $\Jet^1\rho$ the first-order jet manifold of~$\rho$.
Its elements, called 1-jets,
are equivalence classes of local sections of~$\rho$:
two sections are equivalent at a point when they are tangent.
We denote by $\jet^1_t\xi$ the 1-jet of a section $\xi$ at~$t$.

$\Jet^1\rho$ is a fibre bundle over $M$ and over $\Real$,
with canonical projections shown in the diagram:
$$
\xymatrix{
 {\Jet^1\rho} \ar[d]_{\rho_{1,0}} \ar[dr]^{\rho_1} \\
 M \ar[r]_{\rho} & \Real }
$$
We denote by $t$ the canonical coordinate of~$\Real$.
If $(t,q^i)$ are fibred coordinates on $M$,
then $\Jet^1\rho$ has induced coordinates
$(t,q^i,v^i)$.

The bundle $\rho_{1,0} \colon \Jet^1\rho \to M$
is an affine bundle modelled on the vertical bundle of~$\rho$, 
$\Ver\rho\to M$.
Recall that the vertical bundle $\Ver\rho$ is
a subbundle of $\Tan M$,
given by $\Ver\rho = \Ker \Tan \rho$.
The elements of $\Ver\rho$
are the tangent vectors of $M$ which are tangent to the fibres;
equivalently,
$\Ver\rho = \{v \in \Tan M \mid i_{v}\dif t=0 \}$.

There is a canonical embedding
$\iota \colon \Jet^1\rho \to \Tan M$, defined as
$\iota(\jet^1_t\xi)=\dot\xi(t)$.
In local coordinates,
$
\iota(t,q^i,v^i)=(t,q^i,1,v^i).
$
Notice that
$\iota(\Jet^1\rho)=\{v \in \Tan M \mid i_{v}\dif t=1\}$.

\subsection{Implicit systems}

In general, a (time-dependent) 
{\it implicit differential equation}
is defined by a submanifold
$\calD \subset \Jet^1\rho$.
A local section $\xi \colon I \to M$ of~$\rho$ is a
{\it solution} of the differential equation if
\beq
\jet^1\xi(t) \in \calD
\label{xi-impl}
\eeq
for each~$t$.
If the subset $\calD$ is locally described in coordinates
by some equations
$F^\alpha(t,q^i,v^i)=0$,
then the differential equation reads
$F^\alpha(t,\xi^i(t),\dot \xi^i(t)) = 0$.

Suppose that $\calD$ is the image of a jet field,
that is, of a section
$X \colon M \to \Jet^1\rho$.
Then the solutions of the differential equation are
the integral sections of~$X$,
which are the solutions of the explicit differential equation
$$
\jet^1\xi = X \circ \xi .
$$
In coordinates, if $X(t,q^i) = (t,q^i,X^j(t,q^i))$,
the differential equation reads
$\dot \xi^i(t) = X^i(t,\xi^j(t))$.

Consider again $\calD \subset \Jet^1\rho$.
Given a jet field~$X$,
its integral sections are
solutions of the implicit equation defined by $\calD$
iff
\beq
X(M) \subset \calD .
\label{X-impl}
\eeq
So, in a certain sense, solving this equation is equivalent to
solving the implicit equation~(\ref{xi-impl}).

For an explicit differential equation there always exist
solutions,
and each initial condition $x \in M$ defines a unique maximal
solution.
For an implicit differential equation existence and uniqueness
may fail;
in this case one is lead to study the subset of points covered
by solutions, and the multiplicity of solutions.

\subsection{Linearly singular systems}

A (time-dependent) {\it linearly singular system} on~$M$
is defined by
a vector bundle $\pi\colon E \to M$ and
an affine bundle morphism
${\cal A} \colon \Jet^1\rho \to E$:
\beq
\xymatrix{
{\Jet^1\rho} \ar[r]^{\cal A} \ar[d]_{\rho_{1,0}}
   & E \ar[dl]^{\pi} \\
M }
\eeq
For the sake of brevity,
we will refer to this linearly singular system
simply as~$\cal A$.

The system $\cal A$ has an
associated implicit system
given by
\beq
\calD = {\cal A}^{-1}(0) \subset \Jet^1\rho .
\eeq
A local section $\xi \colon I \to M$ is a solution of~$\calD$, 
equation~(\ref{xi-impl}), 
iff it is a solution of the
linearly singular differential equation
\beq
{\cal A} \circ \jet^1\xi = 0 .
\label{xi-sls}
\eeq

In local coordinates,
the bundle morphisms are given by
$$
\pi (t,q^i,u^\alpha) = (t,q^i) ,
\qquad
{\cal A} (t,q^i,v^i) = 
(t,q^i,{\cal A}^\alpha_j(t,q^i)v^j + c^\alpha(t,q^i)) ,
$$
thus the differential equation reads
$$
{\cal A}^\alpha_j(t,\xi^i(t)) \,\dot\xi^j(t) + c^\alpha(t,\xi^i(t)) = 0 .
$$

As before,
it may be convenient to describe the solutions of the
differential equation as integral sections of jet fields.
A jet field $X \colon M \to \Jet^1\rho$ is a
solution jet field of $\cal D$, equation~(\ref{X-impl}),
iff 
\beq
{\cal A} \circ X=0.
\label{X-sls}
\eeq
Then its integral sections are solutions of the differential
equation defined by~$\cal A$.

Locally, $X(t,q^i)=(t,q^i,X^i(t,q^i))$ is a
solution jet field of $\cal A$ when
$$
{\cal A}^\alpha_j(t,q^i) X^j(t,q^i) + c^\alpha(t,q^i) = 0 .
$$

Let us remark that, 
instead of a vector bundle, 
we could have considered an affine bundle $E$ with a section~$b$ 
and an affine bundle morphism ${\cal A} \colon \Jet^1\rho \to E$. 
This slight generalization does not seem too relevant for applications, 
and indeed the section~$b$ endows $E$ with a vector bundle structure.

\section{Constraint algorithm}

In general,
an implicit system does not have solution jet fields,
and does not have solution sections passing through
every point in~$M$.
We want to find a maximal subbundle
$\rho'\colon M' \to \Real$ of~$\rho$
(over $\Real$ for simplicity, 
but more general situations could occur)
where there exist
solution jet fields 
$X \colon M' \to \Jet^1\rho'$ and
solution sections 
$\xi \colon I \to M'$ through every point in~$M'$.

To this end, we can adapt the constraint algorithms
of the time-independent case,
both for implicit systems
\cite{RR-94}
\cite{MMT-95}
and linearly singular systems
\cite{GP-unif,GP-gener},
to the time-dependent case.
A constraint algorithm for 
a time-dependent implicit equation in a product 
$M = \Real \times Q$
has been recently discussed in
\cite{Del-tesi}.

\subsection{Implicit systems}

Let $\calD \subset \Jet^1\rho$ be an implicit system.
We say that a $1$-jet $y \in \calD$ is
{\it integrable} (or {\it locally solvable}) if
there exists a solution $\xi \colon I \to M$ of $\calD$
such that
$\jet^1\xi$ passes through $y$.
One of the purposes of the constraint algorithm 
is to find the set $\calD_{\rm int}$ of all
integrable $1$-jets.

If a solution passes through a point $x \in M$,
necessarily $x$ belongs to the subset
\beq
M_{(1)}:=\rho_{1,0}(\calD) .
\eeq
Denote by
$\rho_{(1)} \colon M_{(1)} \to \Real$
the restriction of~$\rho$ to~$M_{(1)}$.
To proceed with the algorithm,  
we will assume that $\rho_{(1)}$ is a subbundle of~$\rho$.
In this case,
the inclusion $i_{(1)} \colon M_{(1)} \hookrightarrow M$
has a 1-jet prolongation,
$\jet^1 i_{(1)} \colon \Jet^1\rho_{(1)} \hookrightarrow \Jet^1\rho$.
By means of this inclusion, we can define
\beq
\calD_{(1)} := \Jet^1\rho_{(1)} \cap \calD .
\label{D1}
\eeq
Since the solutions of~$\calD$ lay on~$M_{(1)}$,
the integrable jets of~$\calD$ must be contained in $\calD_{(1)}$.
If this is a submanifold,
we have obtained a new implicit system, now on $M_{(1)}$.

This procedure can be iterated:
from $M_{(0)} = M$ and $\calD_{(0)} = \calD$,
and assuming that at each step one obtains
subbundles and submanifolds,
one may define
$M_{(i)} := \rho_{1,0}(\calD_{(i-1)})$
and
$\calD_{(i)} := \Jet^1\rho_{(i)} \cap \calD_{(i-1)}$.
The algorithm finishes when,
for some $k$,
we have $M_{(k+1)}=M_{(k)}$.
In this case,
since $\rho_{1,0}(\calD_{(k)})=M_{(k)}$,
if we suppose for instance that the projection
$\calD_{(k)} \to M_{(k)}$ is a submersion,
we have that
$\calD_{\rm int}=\calD_{(k)}$.

\subsection{Linearly singular systems}

Let ${\cal A} \colon \Jet^1 \rho \to E$ be a
time-dependent linearly singular system
as described in section~2.
We can proceed by applying
the preceding algorithm for implicit systems,
and also by adapting the algorithm for time-independent linearly
singular systems.

So we begin with
$\calD = {\cal A}^{-1}(0)$.
As before, the configuration space must be restricted to
$M_{(1)}:=\rho_{1,0}(\calD)$,
which can also be described as
$$
M_{(1)} = \{x \in M \mid 0_x \in \Img {\cal A}_x\} ;
$$
note that $0_x \in \Img {\cal A}_x$ is 
the necessary consistency condition for 
(\ref{X-sls}) to hold on a given point $x \in M$.

As above, we assume that
$\rho_{(1)} \colon M_{(1)} \to \Real$ is a subbundle of~$M$.

Let us restrict all the data to~$M_{(1)}$:
$
{\cal A}_{(1)}:=\restric{{\cal A}}{\Jet^1\rho_{(1)}}
$,
$
E_{(1)}:=\restric{E}{M_{(1)}}
$,
and
$
\pi_{(1)}:=\restric{\pi}{M_{(1)}}
$.
So we obtain a linearly singular system on~$M_{(1)}$:
\beq
\xymatrix{
{\Jet^1\rho_{(1)}} \ar[r]^{{\cal A}_{(1)}} \ar[d]_{(\rho_{(1)})_{1,0}}
   & E_{(1)} \ar[dl]^{{\pi}_{(1)}} \\
M_{(1)} }
\eeq
It is clear that
the implicit system defined by ${\cal A}_{(1)}$
coincides with
the implicit system $\calD_{(1)}$ obtained above,~(\ref{D1}),
that is,
\beq
{\cal A}_{(1)}^{-1}(0) = \Jet^1\rho_{(1)} \cap \calD .
\eeq

Thus,
if we assume for instance that
each $M_{(i)}$ is a subbundle of $M_{(i-1)}$,
we obtain a constraint algorithm for the linearly singular case:
$$
M_{(i)} := \{ x\in M_{(i-1)} \mid 0_x \in \Img ({\cal A}_{(i-1)})_x\} ,
$$
$$
{\cal A}_{(i)} := \restric{{\cal A}_{(i-1)}}{\Jet^1\rho_{(i)}} ,
$$
$$
E_{(i)}:=\restric{E_{(i-1)}}{M_{(i)}} ,
$$
$$
\pi_{(i)}:=\restric{\pi_{(i-1)}}{M_{(i)}} .
$$
When the algorithm finishes, we arrive to a final system
which is integrable everywhere.
$$
\xymatrix{
{\Jet^1\rho_{(k)}} \ar[r]^{{\cal A}_{(k)}} \ar[d]_{(\rho_{(k)})_{1,0}}
   & E_{(k)} \ar[dl]^{{\pi}_{(k)}} \\
M_{(k)} }
$$

\section{From non-autonomous to autonomous systems}
\label{sec-from}

It is usual to convert a time-dependent system into a 
time-independent one 
by considering the evolution parameter as an new 
dependent variable. 
From a geometric viewpoint, 
this is easily done with an implicit system 
$D \subset \Jet^1\rho$ 
by means of the canonical inclusion 
$\iota \colon \Jet^1\rho \to \Tan M$: 
its image $E = \iota(D)$ 
is a submanifold of $\Tan M$, 
so it defines an autonomous implicit equation. 
The equivalence between both equations is immediate:
\begin{prop}
\label{eq-impl}
A map $\xi \colon \Real \to M$
is a solution section of the time-dependent system $D$
iff
it is a solution path of the autonomous system $E$ such that
$\rho(\xi(t_0))=t_0$ for any arbitrarily given~$t_0$.
\qed
\end{prop}

Now let us focus on the case of 
a time-dependent linearly singular system~$\cal A$:
\beq
\xymatrix{
{\Jet^1\rho} \ar[r]^{\cal A} \ar[d]_{\rho_{1,0}}
   & E \ar[dl]^{\pi} \\
M }
\label{lss}
\eeq
We will also relate this system to an autonomous one.
The main motivation for finding such a relation is that 
the constraint algorithm described in the preceding section 
is easier to implement in the autonomous case.
The reason is that, 
instead of jet fields, 
vector fields can be used to obtain constraint functions 
that define the submanifolds in the constraint algorithm, 
as will be shown later on.

Two constructions to achieve our goal will be proposed.
In the first one, we use
a connection to define a complement of $\Ver\rho$ in $\Tan M$.
In the second construction, we use the vector hull functor 
described in the appendix 
to define vector bundles and morphisms 
from affine bundles and morphisms.

\medskip

Previously, we shall recall some definitions concerning
the autonomous case
\cite{GP-gener}.
An {\it autonomous linearly singular system} on a
manifold $N$ is defined by a vector bundle
$\pi\colon F \to N$, a vector bundle morphism
$A\colon\Tan N \to F$, and a section
$b\colon N \to F$:
$$
\xymatrix{
{\Tan N} \ar[r]^A \ar[d]_{\tau_N}
   & F \ar@<-.5ex>[dl]_{\pi} \\
N \ar@<-.5ex>[ur]_b }
$$
Taking local coordinates $(x^i,\dot x^i)$ on $\Tan N$
and $(x^i,u^\alpha)$ on $F$,
the local expressions of the maps are
$$
\pi(x^i,u^\alpha)=(x^i) ,
\qquad
b(x^i)=(x^i,b^\alpha(x^i)) ,
\qquad
A(x^i,\dot x^i)=(x^i,A^\alpha_j(x^i)\dot x^j) .
$$
We say that a path $\gamma\colon I\to N$ is a
{\it solution path} if
$$
A\circ\dot\gamma=b\circ\gamma .
$$
Locally,
$
A^\alpha_j(\gamma(t))\dot\gamma^j(t)=b^\alpha(\gamma(t)) 
$.
A vector field $X \in \Vect(N)$
is a {\it solution vector field} when
$$
A\circ X=b ;
$$
in coordinates,
$
A^\alpha_j(x)X^j(x)=b^\alpha(x) 
$.

Let us roughly recall
---see \cite{GP-gener} for a detailed description---
how the constraint
algorithm is explicitly carried out in this autonomous
case, that is, how the consecutive constraint manifolds
are effectively computed.
First of all, it can be seen that
the primary constraint
submanifold $M_1=\{x\in M \mid b(x)\in\Img A_x\}$
is locally described by the vanishing of the
functions $\phi^\alpha := \langle s^\alpha, b\rangle$,
where~$(s^\alpha)$ is a local frame
for $\Ker \transp{A}\subset F^*$. 
The constancy of the rank of~$A$ is needed here to ensure 
that this procedure works. 

We have that a vector field~$X$ in~$M$, in order to be
a solution of the system, must satisfy
the equation $A\circ X \feble{M_1} b$.
Vector fields satisfying
this condition always exist and are called primary
vector fields. Given one primary vector field~$X_0$,
the others have the form
$X \feble{M_1} X_0 + \sum_\mu f^\mu\Gamma_\mu$,
where~$f^\mu$ are functions uniquely determined
on~$M_1$ and~$(\Gamma_\mu)$ is a local frame
for~$\Ker A$.

A primary vector field~$X$ can be a solution of the
system only if it is tangent to~$M_1$, so we obtain
the equation, for every constraint~$\phi^\alpha$,
$(X\cdot\phi^\alpha)(x) = 0$, for~$x\in M_1$, or,
equivalently,
$(X_0\cdot\phi^\alpha)(x)+
\sum_\mu (\Gamma_\mu\cdot\phi^\alpha)(x) f^\mu(x) = 0$.
These equations may provide new constraints that define
the secondary constraint submanifold~$M_2$, 
and may also determine some of the functions~$f^\mu$.

This procedure can be iterated until we determine
which primary vector
fields are solutions of the system,
and we obtain the final
submanifold where they are defined.

\subsection{Jet field construction}

Consider the linearly singular system given by~(\ref{lss}).
Let us choose an arbitrary jet field $\Gamma\colon M\to\Jet^1\rho$.
This jet field $\Gamma$ induces 
\cite{Sau-jets}
in a natural way a connection $\tilde\Gamma$
on the bundle $\rho$ and a
splitting $\Tan M=\Ver\rho\oplus H_\Gamma$
of the tangent bundle, with
projections $v_\Gamma$ and $h_\Gamma$.
The coordinate expressions are:
$$
\Gamma(t,q^i)=(t,q^i,\Gamma^i(t,q^i)) ,
$$
$$
\tilde\Gamma=\dif t\otimes \left(\tanvec{t}+\Gamma^i\tanvec{q^i}\right) ,
$$
$$
v_\Gamma(t,q^i;\dot t,\dot q^i)=(t,q^i;0,\dot q^i - \dot t\,\Gamma^i(t,q^i)) ,
\qquad
h_\Gamma(t,q^i;\dot t,\dot q^i)=(t,q^i;\dot t, \dot t\,\Gamma^i(t,q^i)) .
$$
From this we can define a section of~$\pi$,
$$
b_\Gamma:=-{\cal A}\circ\Gamma\colon M\to E,
$$
and a vector bundle morphism,
$$
A_\Gamma:=\vec{\cal A}\circ v_\Gamma\colon\Tan M\to E,
$$
where
$\vec{\cal A} \colon \Ver\rho \to E$ is the vector bundle morphism
associated with the affine map~${\cal A}$.

With these objects we can construct a
time-independent linearly singular system:
\beq
\xymatrix{
{\Tan M} \ar[r]^-{A_\Gamma\oplus\dif t} \ar[d]_{\tau_M}
   & E\oplus\Real \\
M \ar[ur]_{b_\Gamma\oplus 1} }
\label{jlss}
\eeq

This system is equivalent to the time-dependent
system (\ref{lss}) in the sense of the following proposition.

\begin{prop}
\label{eq-jetfield}
Consider the time-dependent system given by (\ref{lss}).
Given any jet field $\Gamma\colon M\to\Jet^1\rho$, we have:
\begin{itemize}
\item[i)]
A map $\xi\colon\Real\to M$
is a solution section of the time-dependent
system (\ref{lss}) if, and only if,
it is a solution path of
the autonomous system (\ref{jlss}) such that
$\rho(\xi(t_0))=t_0$ for any arbitrarily given~$t_0$.
\item[ii)]
A map $X\colon M\to \Jet^1\rho$
is a solution jet field of the time-dependent system (\ref{lss})
if, and only if,
considered as a vector field in~$M$,
it is a solution vector field of
the autonomous system~(\ref{jlss}).
\end{itemize}
\end{prop}

\noindent
(Note that we are using the embedding $\iota\colon\Jet^1\rho\to\Tan M$
defined in section~$2$ to identify jet fields as vector fields.)

\proof
It is immediate, taking into account the local expressions of
the equations defined respectively by both systems:
\begin{itemize}
\item
${\cal A}^\alpha_j(t,q^i)\,v^j+c^\alpha(t,q^i)=0$
\item
$\left\{
\begin{array}{l}
{\cal A}^\alpha_j(t,q^i)\,(\dot q^j-\dot t\,\Gamma^j(t,q^i))=
-{\cal A}^\alpha_j(t,q^i)\,\Gamma^j(t,q^i)-c^\alpha(t,q^i)\\
\dot t=1
\end{array}
\right.$
\qed
\end{itemize}

\subsection{Vector hull construction}
\label{ssec-hull}

Here we will apply the vector hull functor described in the appendix. 
The affine bundle morphism $\cal A$ in (\ref{lss}) induces
a vector bundle morphism $\widehat{\cal A}$ between
the vector hulls of $\Jet^1\rho$ and $E$:
$$
\xymatrix{
 *++{\Jet^1\rho} \ar@<-.5ex>@{^{(}->}[r] \ar[d]_{\cal A}
   & \widehat{\Jet^1\rho} \ar[d]^{\widehat{\cal A}} \\
  *++{E} \ar@<-.5ex>@{^{(}->}[r]_i
   & \widehat E }
$$
The $0$ section of $\pi \colon E \to M$ also induces a section
$\hat{0}$ of $\widehat{\pi} \colon \widehat{E} \to M$, 
defined by $\hat{0}=i\circ 0$.
Recall that with the identification 
$\widehat{E} = E \times \Real$, 
we have $\hat 0 = (0,1)$. 

Using the canonical identification of $\widehat{\Jet^1\rho}$ with
$\Tan M$, we can construct the following
linearly singular system:
\beq
\xymatrix{
 {\Tan M} \ar[r]^{\widehat{\cal A}} \ar[d]_{\tau_M} & \widehat{E} \\
 M \ar[ur]_{\hat{0}}}
\label{hlss}
\eeq

This system is equivalent to the time-dependent system (\ref{lss}) 
in the sense of the following proposition.

\begin{prop}
\label{eq-vhull}
Consider the time-dependent system given by (\ref{lss}).
\begin{itemize}
\item[i)]
A map $\xi\colon\Real\to M$
is a solution section of the time-dependent
system (\ref{lss}) if, and only if,
it is a solution path of
the associated autonomous system (\ref{hlss}) such that
$\rho(\xi(t_0))=t_0$ for any arbitrarily given~$t_0$.
\item[ii)]
A map $X\colon M\to \Jet^1\rho$
is a solution jet field of the time-dependent system (\ref{lss})
if, and only if,
considered as a vector field in~$M$,
it is a solution vector field of
the associated autonomous system~(\ref{hlss}).
\end{itemize}
\end{prop}

\proof
Again we can prove the result in local coordinates,
where the equations of the systems
(\ref{lss}) and (\ref{hlss}) read, respectively,
\begin{itemize}
\item
$c^\alpha(t,q^i)+{\cal A}^\alpha_j(t,q^i)v^j=0$
\vskip 1mm
\item
$\left\{
\begin{array}{l}
c^\alpha(t,q^i)\dot t+{\cal A}^\alpha_j(t,q^i)\dot q^j=0
\\
\dot t=1
\end{array}
\right.$
\end{itemize}

We have used the coordinates $(u^0,u^\alpha)$ on $\widehat{E}$ induced
by the affine frame $(0;u^\alpha)$ on~$E$.
\qed

\section{The second-order case}

The preceding results can be extended to consider
higher-order implicit and linearly singular
differential equations.
Of course, the most important case,
due to its applications to mechanics,
is that of second-order equations,
to which we devote this section.
Before proceeding, 
we need some additional facts about the second-order
jet bundle of a fibre bundle
$\rho \colon M \to \Real$.

\subsection{Second-order jet bundles}

The jet space $\Jet^2\rho$ is a fibre bundle over $\Jet^1\rho$,
$M$ and~$\Real$.
The canonical projections are:
$$
\xymatrix{
 {\Jet^2\rho} \ar[r]_{\rho_{2,1}} \ar@/_/[rrd]_{\rho_2}
              \ar@/^1pc/[rr]^{\rho_{2,0}}
  & {\Jet^1\rho} \ar[r]_{\rho_{1,0}} \ar[rd]_{\rho_1}
    & M \ar[d]^{\rho} \\
  & & \Real }
$$
The bundle $\rho_{2,1}\colon\Jet^2\rho\to\Jet^1\rho$ is
an affine bundle modelled on the vertical bundle of~$\rho_{1,0}$, 
$\Ver\rho_{1,0}\to\Jet^1\rho$.

There exists a natural embedding
$\jmath\colon\Jet^2\rho\to\Jet^1\rho_1$, defined by
$\jmath(\jet^2_t\xi)=\jet^1_t(\jet^1\xi)$.
In addition to~$\jmath$,
we can use the embedding
$\iota \colon \Jet^1\rho_1 \to \Tan\Jet^1\rho$
to include the second-order jet bundle into a tangent bundle:
$$
\begin{array}[t]{ccccc}
\Jet^2\rho & \stackrel{\jmath}{\longrightarrow} & \Jet^1\rho_1 & \stackrel{\imath}{\longrightarrow} & \Tan\Jet^1\rho \\
\jet^2_t\xi & \longrightarrow & \jet^1_t(\jet^1\xi) & \longrightarrow &
(\jet^1\xi)^{{\textstyle .}}(t)
\end{array}
$$

Consider natural coordinates $(t,q^i,v^i,a^i)$ on $\Jet^2\rho$.
With them,
the local expression of this composed embedding 
$\kappa = \iota\circ\jmath$ 
reads
$$
\kappa(t,q^i,v^i,a^i)=(t,q^i,v^i,1,v^i,a^i).
$$
This shows that the image of $\Jet^2\rho$ by the embedding
can be expressed as
$$
\kappa(\Jet^2\rho) =
\{ w \in \Tan\Jet^1\rho \mid
i_{w}\dif t=1, \; \vend(w)=0 \} .
$$
Here there appears another relevant operator,
the canonical vertical endomorphism
$\vend$ of $\Tan\Jet^1\rho$,
whose local expression is
$$
\vend=(\dif q^i-v^i \dif t)\otimes\tanvec{v^i} ;
$$
note also that
$\Img(\vend)=\Ver\rho_{1,0}$.

The Cartan distribution of~$\Jet^1\rho$ is
the kernel of the vertical endomorphism~$\vend$
of~$\Tan(\Jet^1\rho)$.
We denote it by~$C\rho_{1,0}$.
Locally, we can describe~$C\rho_{1,0}$
as the distribution generated by the
$n+1$ vector fields
$\{\tanvec{t}+ v^i\tanvec{q^i},
\tanvec{v^i}\}$.
We have an exact sequence
\beq
\xymatrix{
  0 \ar[r] 
& C\rho_{1,0}\; \ar[r]
& \Tan(\Jet^1\rho) \ar[r]^-{\vend}
& \Ver\rho_{1,0} \ar[r] 
& 0
}
\label{S-exact}
\eeq

\subsection{Second-order implicit and linearly singular systems}

Similar to the first-order case,
a {\it second-order implicit differential equation}
is defined by a submanifold
$\calD \subset \Jet^2\rho$.
A local section $\xi \colon I \to M$ is a
{\it solution} of the differential equation if
$
\jet^2\xi(t) \in \calD
$
for each~$t$.
In coordinates, this equation can be expressed as
$F^\alpha(t,\xi^i(t),\dot \xi^i(t),\ddot \xi^i(t)) = 0$.
Like in the first-order case,
if $\calD$ is the image of a section
$X \colon \Jet^1\rho \to \Jet^2\rho$,
the equation can be written in normal form.

Now let us consider the linearly singular case.
A {\it time-dependent second-order linearly singular system}
is defined by
a vector bundle $\pi\colon E\to\Jet^1\rho$ and
an affine bundle morphism
${\cal A} \colon \Jet^2\rho \to E$:
\beq
\xymatrix{
{\Jet^2\rho} \ar[r]^{\cal A} \ar[d]_{\rho_{2,1}}
   & E \ar[dl]^{\pi} \\
{\Jet^1\rho} }
\label{2lss}
\eeq
Its {\it solution sections} are sections
$\xi$ of~$\rho$ such that
$$
{\cal A}\circ\jet^2\xi=0 .
$$
Locally this reads
\beq
{\cal A}^\alpha_j(t,\xi^i(t),\dot\xi^i(t)) \,\ddot\xi^j(t)
+c^\alpha(t,\xi^i(t),\dot\xi^i(t))=0.
\label{loc2lss}
\eeq
A second-order jet field,
that is,
a section $X$ of~$\rho_{(2,1)} \colon \Jet^2\rho \to \Jet^1\rho$,
is a {\it solution jet field} if
$$
{\cal A}\circ X=0 ;
$$
locally this reads
$
{\cal A}^\alpha_j(t,q^i,v^i)X^j(t,q^i,v^i)+c^\alpha(t,q^i,v^i)=0 
$.

In a similar way of what we did in the previous section,
it is interesting to convert the singular system given by
(\ref{2lss})
into a first-order autonomous linearly singular system.
As before, we present two constructions of this.

\subsection{Jet field construction}

Choose an arbitrary second-order jet field
$\Gamma\colon\Jet^1\rho\to\Jet^2\rho$.
Again this determines a splitting of the tangent bundle 
of $\Jet^1\rho$ as a direct sum
$\Tan\Jet^1\rho = \Ver\rho_{1,0} \oplus H_\Gamma$, 
with associated projections $v_\Gamma$ and~$h_\Gamma$.
Now we define a section of~$\pi$ 
$$
b_\Gamma:=-{\cal A}\circ\Gamma\colon\Jet^1\rho\to E
$$
and a vector bundle morphism
$$
A_\Gamma:=\vec{\cal A}\circ v_\Gamma\colon\Tan\Jet^1\rho\to E.
$$

With these definitions,
we obtain an autonomous first-order linearly singular system
on the manifold $\Jet^1\rho$:
\beq
\xymatrix{
 {\Tan \Jet^1\rho} \ar[d] \ar[rr]^-{A_\Gamma\oplus\vend\oplus\dif t}
  & & E \oplus \Ver \rho_{1,0} \oplus \Real \\
 {\Jet^1\rho} \ar[urr]_-{b_\Gamma\oplus 0\oplus 1} }
\label{j2lss}
\eeq

A result quite similar to Proposition~\ref{eq-jetfield} 
can be formulated,
in the sense that this system is equivalent to
the original time-dependent second-order system~(\ref{2lss}).
This can be readily seen by comparing 
the local expression~(\ref{loc2lss}) 
with that of the equation defined by system~(\ref{j2lss}).
We skip the details. 

\subsection{Vector hull construction}

As opposite to the first-order case,
the vector hull of $\Jet^2\rho$
can not be identified with a tangent bundle,
but rather with a tangent subbundle.
As it is shown in the appendix,
$\widehat{\Jet^2\rho}$
can be identified with the Cartan distribution $C\rho_{1,0}$
on $\Jet^1\rho$.
Then, as in section~\ref{ssec-hull},
if we homogenize the system (\ref{2lss})
we obtain the following:
\beq
\xymatrix{
 C\rho_{1,0} \ar[r]^{\widehat{\cal A}} \ar[d]_{\tau_{\Jet^1\rho}}
   & {\widehat{E}}  \\
 {\Jet^1\rho} \ar[ur]_{\hat{0}}}
\label{h2lss}
\eeq
This is an autonomous linearly singular system on $\Jet^1\rho$,
except for the fact that
there is only a subbundle $C\rho_{1,0} \subset \Tan\Jet^1\rho$ 
instead of the whole tangent bundle.
The interpretation of this system is the same as in the ordinary
case, but with the additional requirement that,
for a path $\eta \colon I \to \Jet^1\rho$,
its derivative $\dot\eta$ must lie in $C\rho_{1,0}$
---which is a natural condition if $\eta$ has to be the lift
$\jet^1\xi$ of a section of~$\rho$. 
In coordinates, if $\eta = (t, q^i, v^i)$, 
this requirement amounts to
$$
\dot q^i = \dot t \,v^i.
$$
Assuming that $\dot\eta\subset C\rho_{1,0}$, the local equations
for the path $\eta$ to be a solution of
the system~($\ref{h2lss}$) are
$$
\left\{
\begin{array}{l}
\dot t = 1 \\
\dot t \,c^\alpha(t,q^i,v^i) + \dot v^j A_j^\alpha(t,q^i,v^i) = 0
\end{array}
\right.
$$
Comparing these three equations 
with equation~(\ref{loc2lss}), 
we see that 
systems~(\ref{2lss}) and~(\ref{h2lss}) are equivalent.

Finally, 
we will show that
the system~(\ref{h2lss}) 
can be expressed as a linearly singular system, 
provided we have an appropriate extension of~${\cal A}$. 
Since $E$ is a vector bundle, 
we have a canonical identification 
$\widehat{E} = E \oplus \Real$, 
and the vector extension $\widehat{\cal A}$ 
can be written 
$\widehat{\cal A} = {\cal A}\hh \oplus \dif t$ 
(see the appendix). 
Suppose that we have an extension 
$\bar A \colon \Tan(\Jet^1\rho) \to E$ 
of the map 
${\cal A}\hh \colon C\rho_{1,0} \to E$; 
in some applications (see next section) 
there exists a natural extension~$\bar A$.
Then the system~(\ref{h2lss}) 
can be described as the following linearly singular system: 
\beq
\xymatrix{
 {\Tan \Jet^1\rho} \ar[d] \ar[rr]^-{\bar A \oplus \dif t \oplus \vend}
  & & E \oplus \Real \oplus \Ver \rho_{1,0} 
\\
 {\Jet^1\rho} \ar[urr]_-{0 \oplus 1 \oplus 0} }
\label{h2lssext}
\eeq
The only thing to be noted is that $C\rho_{1,0}$ 
is the kernel of~$\vend$, 
see~(\ref{S-exact}).

\section{Some applications to mechanics}

\subsection{Non-autonomous lagrangian systems}

Let $\rho\colon Q\to\Real$ be a fibre bundle modelling 
a time-dependent configuration space.
A lagrangian system on $Q$ is determined by a lagrangian function
$L\colon\Jet^1\rho\to\Real$.

The vertical endomorphism $\vend$ of $\Tan(\Jet^1\rho)$ 
allows to construct the Poincar\'e--Cartan forms
$$
\Theta_L= \transp{\vend} \circ \dif L + L \dif t
\,\in\, \Omega^1(\Jet^1\rho) ,
$$
$$
\Omega_L = -\dif\Theta_L 
\,\in\, \Omega^2(\Jet^1\rho) .
$$
By contraction, this one defines a morphism 
$
\hat \Omega_L \colon \Tan(\Jet^1\rho) \to \Tan^*(\Jet^1\rho)
$.

In coordinates $(t,q,v,a)$ of $\Jet^2\rho$, 
the Euler--Lagrange equation can be written 
$\dif \hat p/\dif t = \partial L/\partial q$, 
where $\hat p = \partial L/\partial v$. 
Now, given a vector field
$X \in \Vect(\Jet^1\rho)$, 
we can compute 
$$
i_X \Omega_L = 
\left( (X \lop q)-v(X \lop t) \vphantom{\derpar{}{}} \right) \dif \hat p 
- \left( (X \lop \hat p) - \derpar{L}{q} (X \lop t) \right) \dif q 
+ \left( (X \lop \hat p)v - \derpar{L}{q} (X \lop q) \right) \dif t .
$$
Consider the case where $X$ 
is a second-order vector field,
$
\ds X = \derpar{}{t} + v \derpar{}{q} + A(t,q,v) \derpar{}{v}
$, 
which amounts also to
\beq
i_X \dif t = 1 ,
\qquad
\vend \circ X = 0 .
\label{eq-2nd}
\eeq
Then the preceding expression simplifies to 
$\ds
i_X \Omega_L = 
\left( (X \lop \hat p) - \derpar{L}{q} \right) (v \dif t - \dif q) , 
$
and so the integral curves of~$X$ are a solution of the 
Euler--Lagrange equation iff
\beq
i_X \Omega_L = 0 .
\label{eq-Omega}
\eeq

Now recall the affine inclusion 
$\kappa \colon \Jet^2\rho \hookrightarrow \Tan \Jet^1\rho$, 
which identifies jet fields $\Jet^1\rho \to \Jet^2\rho$ 
with second-order vector fields on $\Jet^1\rho$. 
From the preceding discussion, 
it is clear that the lagrangian dynamics 
may be described by the following 
second-order linearly singular system on~$Q$:
\beq
\xymatrix{
 {\Jet^2\rho} \ar[r]^-{\hat\Omega_L \circ \kappa}
              \ar[d]_{\rho_{2,1}}
 & 
 {\Tan^*\Jet^1\rho} \ar[dl]^{\tau^*_{\Jet^1\rho}} 
\\
 {\Jet^1\rho} }
\eeq
Using the vector hull construction described in the preceding section,  
we can convert this system 
into a first-order autonomous system on $\Jet^1\rho$: 
$$
\xymatrix{
 {\Tan \Jet^1\rho} \ar[d] \ar[rr]^-{\hat\Omega_L \oplus \dif t \oplus \vend}
  & & \Tan^*\Jet^1\rho \oplus \Real \oplus \Ver \rho_{1,0} 
\\
 {\Jet^1\rho} \ar[urr]_-{0 \oplus 1 \oplus 0} }
$$
and note that its equations of motion are precisely 
(\ref{eq-2nd}), (\ref{eq-Omega}). 
If the lagrangian is regular, 
then this linearly singular system is regular;
otherwise, the system is singular 
and the constraint algorithm for linearly singular systems 
can be applied to obtain the dynamics. 

Finally, 
let us note that there are other equivalent descriptions 
of the dynamics. 
For instance, 
instead of $\hat \Omega_L$, 
we could have used the Euler--Lagrange form on $\Jet^2\rho$. 
We omit the details. 


\subsection{Skinner-Rusk formulation of time-dependent mechanics}

A mixed lagrangian-hamiltonian formulation of 
time-independent mechanics 
was studied geometrically in a series of papers by Skinner and 
Rusk 
\cite{Ski-mixt,SR-mixt}. 
Recently, the time-dependent case has been studied in
\cite{CMC-SR}. 
We will show how this can be described in our formalism. 

Our starting point is
a fibre bundle $\rho \colon Q \to \Real$ and a lagrangian
function $L \colon \Jet^1\rho \to \Real$.
In this formulation, the dynamics is represented by a first-order
system on the manifold $M := \Tan^*Q \times_Q \Jet^1\rho$. 
Denote the several projections as in the following diagram:
$$
\xymatrix{
  \llap{$M:=$} \Tan^*Q \times_Q \Jet^1\rho 
     \ar[r]^-{{\rm pr}_2}
     \ar[dr]
     \ar[d]^{{\rm pr}_1}
     \ar `u[r] `r[drr]^{\pi} [drr]
& {\Jet^1\rho} 
      \ar[d]^{\rho_{1,0}}
      \ar[dr]^{\rho_1} 
\\
  {\Tan^*Q} 
      \ar[r]_{\tau^*_Q}
& Q \ar[r]_{\rho}
& \Real 
}
$$
We can define the following function on~$M$:
$$
{\cal H}=
\langle{\rm pr}_1,{\rm pr}_2\rangle-{\rm pr}^*_2 L,
$$
where $\langle \,{,}\, \rangle$ denotes the natural pairing
between vectors and covectors on $Q$, and the 2-form on~$M$
$$
\Omega_{\cal H}={\rm pr}^*_1\omega_Q-\dif {\cal H}\wedge\dif t,
$$
where $\omega_Q$ is the canonical symplectic form on $\Tan^*Q$;
it defines a morphism
$\widehat{\Omega}_{\cal H}\colon\Tan M\to\Tan^* M$.

With these definitions we can write the equations of the dynamics
in the Skinner-Rusk formulation, which are
$$
\left\{
\begin{array}{l}
i_Z\Omega_{\cal H}=0\\
i_Z\dif t= 1\\
\end{array}
\right.
,
$$
for a vector field $Z \in \Vect(M)$.
These equations are equivalent to the
time-dependent linearly singular system on~$M$
defined by the following diagram:
$$
\xymatrix{
 {\Jet^1\pi} \ar[rr]^-{\restric{(\widehat{\Omega}_{\cal H})}{\Jet^1\pi}}
             \ar[d]_{\pi_{1,0}}
  & & {\Tan^* M} \ar[dll]^{\tau^*_M} \\
 M }
$$

\section{An example: a simple pendulum of given variable length}

Consider a simple pendulum whose length is given by
a time-dependent function $R(t)$.
Its equation of motion can be written as
\cite{GP-gener}
$$
\left\{
\begin{array}{ccl}
\dot x & = & v_x \\
\dot y & = & v_y \\
\dot{v_x} & = & -\tau x\\
\dot{v_y} & = & -\tau y - g\\
x^2+y^2 & = & R^2(t)
\end{array}
\right.
,
$$
where $g$ is the gravitational acceleration and $\tau R(t)$
is the string tension per unit mass.

This system can be described as
a time-dependent linearly singular system in the following way.
Take $M:= \Real^6$,
with coordinates $(t,x,y,v_x,v_y,\tau)$,
as a configuration manifold fibred over $\Real$, with coordinate~$t$.
The product $M\times\Real^5$
is a trivial vector bundle over $M$, 
and the affine bundle morphism
${\cal A}\colon\Jet^1\rho\to M\times\Real^5$ defined by
$$
{\cal A}(\dot x,\dot y,\dot{v_x},\dot{v_y},\dot\tau)_p = (\dot
x-v_x,\dot y-v_y,\dot{v_x}+\tau x,\dot{v_y}+\tau
y+g,-(x^2+y^2-R^2(t)))_p,
$$
where $p=(t,x,y,v_x,v_y,\tau)\in M,$
models the system.

Choosing the connection $\tilde\Gamma = \dif t\oplus\tanvec{t}$,
as described in section~\ref{sec-from}, 
we can convert this system into an autonomous 
linearly singular system on~$M$, 
which can be written as
$$
\left(
\begin{array}{cccccc}
0&1&0&0&0&0\\
0&0&1&0&0&0\\
0&0&0&1&0&0\\
0&0&0&0&1&0\\
0&0&0&0&0&0\\
1&0&0&0&0&0
\end{array}
\right)
\left(
\begin{array}{c}
\dot t \\ \dot x \\ \dot y \\ \dot{v_x} \\ \dot{v_y} \\ \dot\tau
\end{array}
\right)
=
\left(
\begin{array}{c}
v_x \\ v_y \\ -\tau x \\ -\tau y -g \\ x^2+y^2-R^2(t) \\ 1
\end{array}
\right)
$$

We solve this autonomous linearly singular system
by means of the constraint algorithm for
the autonomous case 
that is sketched in section~4. In this case,
three steps are needed to solve the system.
We give here only the
constraint functions~$\phi^i$ and
manifolds~$M_i$ obtained at each step:
\begin{enumerate}
\item
$\phi^1=x^2+y^2-R^2(t)$,
\\
$M_1=\{\phi^1=0\}$,
\\
and the possible solution vector fields are of the form
$$X \feble{M_1} \tanvec{t} + v_x\tanvec{x} + v_y\tanvec{y}
-\tau x\tanvec{v_x} - (\tau y +g)\tanvec{v_y}+ f\tanvec{\tau},$$
where $f\in\Cinfty(M_1)$ is a function to be determined.
\item
$\phi^2 = X\cdot\phi^1 \feble{M_1} x v_x + y v_y - RR'$,
\\
$M_2 = \{\phi^1 = \phi^2 = 0 \}$.
\item
$\phi^3 = X\cdot\phi^2 \feble{M_2} v_x^2 + v_y^2 - \tau R^2 - (RR''+(R')^2)$,
\\
$M_3 = \{\phi^1 = \phi^2 = \phi^3 = 0 \}$.
\item
$\phi^4 = X\cdot\phi^3 \feble{M_3} -4\tau RR' - 3v_y g - (RR''' + 3R'R'')- fR^2$.
\\
The equation $\phi^4 = 0$ determines the function $f\in\Cinfty(M_3)$,
so it is not a new constraint and the system is solved.
\end{enumerate}

Using polar coordinates $(t,r,\varphi,v_r,v_\varphi,\tau)$, defined by
$$
\left\{
\begin{array}{ccl}
x & = & r\cos\varphi\\
y & = & r\sin\varphi\\
v_x & = & v_r\cos\varphi - v_\varphi r\sin\varphi\\
v_y & = & v_r\sin\varphi + v_\varphi r\cos\varphi\\
\end{array}
\right.
,
$$
we see that the final submanifold $M_3$ is diffeomorphic to
$\Real\times \Tan {\bf S}^1$,
and it is embedded in $M$ by
$(t,\varphi,v_\varphi)
\to
(t,R(t),\varphi,R'(t),(v_\varphi^2 R(t)-g\sin\varphi -R''(t))/R(t))
$.
The (unique) solution vector field is
$$
X = \tanvec{t} + v_\varphi\tanvec{\varphi}
-\frac{2R'(t)v_\varphi+g\cos\varphi}{R(t)}\tanvec{v_\varphi} .
$$

\appendix
\section*{Appendix: vector hulls}

\subsection*{Vector hulls of affine spaces}

Though affine geometry is well-known,
it presents some less known canonical structures.
The basic fact is that any affine space has a canonical immersion,
as a hyperplane, in a vector space;
with it, affine maps can be understood as linear maps.
We will not go into the details of this construction,
nor into its several applications
---see for instance
\cite{BB-tors,Ber,GM-vhulls,MMS-affbun},
instead we will describe some of the basic facts.

If $A$ is a real affine space,
we denote by $\vec A$ its associated vector space.
We denote by $\Aff(A,B)$ the set of affine maps between
the affine spaces $A$ and~$B$.
For any affine map $f \colon A \to B$ there is an associated
linear map $\vec f \in \Lin(\vec A,\vec B)$.

The \textit{vector hull} of $A$ is
a vector space $\widehat A$ together with an affine map
$j \colon A \to \widehat A$, and such that 
the following universal property holds: 
for every vector space $F$ and affine map $h \colon A \to F$,
there exists a unique linear map $h\hh \colon \widehat A \to F$
such that $h = h\hh \circ j$.
It turns out that 
$j$ is an affine immersion,
with $j(A)$ a hyperplane in~$\widehat A$ not containing~0.
$$
\xymatrix{
*++{A} \ar@{^{(}->}[r]^j \ar[dr]_h  &  \widehat A \ar@{-->}[d]^{h\hh} \\
 & F}
$$
Notice that such a hyperplane can be described as
the set $w^{-1}(1)$,
with $w \colon \widehat A \to \R$ a unique linear form.
This also identifies
$\vec A$ with $w^{-1}(0)$.
We can gather all this information in a diagram:
$$
\xymatrix{
0 \ar[r] & \vec A \ar[r]^{i} & \widehat A \ar[r]^{w} & \R \ar[r] & 0 \\
         &                   & *++{A\;} \ar@{^{(}->}[u]_{j}
}
$$
From the universal property,
it is clear that the assignment $h \mapsto h\hh$
is indeed an isomorphism
$\Aff(A,F) \simeq \Lin(\widehat A,F)$,
and in particular we have
$\Aff(A,\R) \simeq {\widehat A}^*$.

Given an affine map $f \colon A \to B$,
there is a unique linear map
$\widehat f \colon \widehat A \to \widehat B$
such that $\widehat f \circ j_A = j_B \circ f$.
We will call it the \textit{vector extension} of~$f$
for obvious reasons.
$$
\xymatrix{
*++{A} \ar@{^{(}->}[d]_{j_A} \ar[r]^f      & *++{B} \ar@{^{(}->}[d]^{j_B}
\\
\widehat A              \ar@{-->}[r]^{\widehat f}  & \widehat B
}
$$
Now the assignment $f \mapsto \widehat f$
is an affine inclusion
$\Aff(A,B) \hookrightarrow \Lin(\widehat A,\widehat B)$,

\smallskip

Now let us describe the particular, but important,
case where where the affine space is a vector space~$F$.
Then we have a canonical identification $\widehat F = \R \times F$,
with the inclusion $j_F(u) = (1,u)$,
and $w_F(\lambda,u) = \lambda$.
Given an affine map $h \colon A \to F$,
its vector extension
$\widehat{h} \colon \widehat{A} \to \widehat{F} = \Real \times F$
is given by $\widehat{h} = (w_A,h\hh)$. 

More particularly, 
for an affine map $h \colon E \to F$ between vector spaces, 
$h(u) = h_0 + h_1 \cdot u$,
we have that $h\hh \colon E \times \Real \to F$ is given by 
$h\hh(\lambda,u) = h_0 \lambda + h_1 \cdot u$, 
and the vector extension of~$h$ is 
$\widehat{h}(\lambda,u) = (\lambda,h_0 \lambda + h_1 \cdot u)$. 

\smallskip

Finally, let us put coordinates everywhere.
Consider a point $e_0=a_0 \in A$ and 
a basis $(e_i)_{i \in I}$ of~$\vec A$.
Then, with the appropriate identifications,
every point in $\widehat A$ can be uniquely written as
$x = x^0 e_0 + x^i e_i$.
The point $x$ belongs to $A$ iff $x^0=1$, and belongs to $\vec A$ iff $x^0=0$.
With these coordinates,
the vector extension of an affine map 
$y^j = c^j + T^j_{\,i} x^i$
is the linear map
$y^\nu = T^\nu_{\,\mu} x^\mu$,
with $T^j_{\,0} = c^j$, $T^0_{\,0} = 1$, $T^0_{\,i} = 0$.

\subsection*{Vector hulls of affine bundles and jet bundles}

All that we have done up to now with affine spaces can be
formulated for affine bundles in an analogous way
\cite{MMS-affbun}.
In this case, the starting point is
an affine bundle $A\to M$ modelled on
a vector bundle $\vec A \to M$.
Without going into technical details, the vector
hull of~$A$ is the vector bundle $\widehat A \to M$ whose fibres
are the vector hulls of the fibres of~$A$ (which are affine spaces).
All the affine and linear maps that we considered before are
now affine and vector bundle morphisms over the
identity map on~$M$.

For our discussion it is particularly useful
the following fact: if we have an exact
sequence of vector bundle morphisms
$
\xymatrix{
0 \ar[r] & **[u]{\vec A} \ar[r]^-\alpha & W \ar[r]^- w
 & {M\times\Real} \ar[r] & 0
}
$,
then $w^{-1}(1)$ is an affine bundle modelled
on~$\vec A$ (so it is isomorphic to~$A$),
and~$W$ is canonically isomorphic to
the vector hull of $w^{-1}(1)$.

We will apply this to find the vector hulls of the
affine bundles that play a role in this paper, that is,
the jet bundles $\rho_{1,0} \colon \Jet^1\rho \to M$
and $\rho_{2,1}\colon\Jet^2\rho\to\Jet^1\rho$,
when $\rho\colon M\to \Real$ is a fibre bundle
over the real line.
Recall sections~2.1 and~5.1 for the definition
and properties of these jet bundles.

From section~2.1 it follows that
the sequence of vector bundle morphisms
$$
\xymatrix{
0 \ar[r] & \Ver\rho\; \ar@{^{(}->}[r] & \Tan M \ar[r]^-{\dif t}
 & {M\times\Real} \ar[r] & 0
}
$$
is exact, where~$\dif t$ denotes, by
abuse of notation, the contraction of a tangent vector with the
1-form~$\dif t$. Furthermore, it is also seen in section~2.1
that~$\Jet^1\rho$ can be canonically embedded
into~$\Tan M$, and the image of this embedding
is just~$\dif t^{-1}(1)$.
Therefore, the vector hull of~$\Jet^1\rho$ is
naturally identified with~$\Tan M$:
$$
\widehat{\Jet^{1}\rho}=\Tan M.
$$

Now we consider
the bundle $\rho_{2,1}\colon\Jet^2\rho\to\Jet^1\rho$.
We have seen in section~5.1 that
it is an affine bundle modelled
on the vector bundle~$\Ver\rho_{1,0}$,
which is a subbundle of
the tangent bundle~$\Tan(\Jet^1\rho)$, and that
$\Jet^2\rho$ can be naturally embedded
into~$\Tan(\Jet^1\rho)$.
Our aim is to identify
the vector hull~$\widehat{\Jet^2\rho}$
with a suitable subbundle of~$\Tan(\Jet^1\rho)$.

The Cartan distribution 
$C\rho_{1,0}$ is a subbundle of~$\Tan(\Jet^1\rho)$ which includes
$\Ver\rho_{1,0}$, and it is straightforward to see that the sequence
of vector bundle morphisms
$$
\xymatrix{
0 \ar[r] & \Ver\rho_{1,0}\; \ar@{^{(}->}[r]
 & C\rho_{1,0} \ar[r]^-{w}
 & {\Jet^1\rho\times\Real} \ar[r] & 0
}
$$
is exact, where~$w$ denotes denotes the contraction
with the 1-form~$\dif t$, restricted to vectors
in~$C\rho_{1,0}$.
It is easy to see that~$w^{-1}(1)$
is equal to~$\Jet^2\rho$ as a submanifold
of~$\Tan(\Jet^1\rho)$, so
we can conclude that~$\widehat{\Jet^2\rho}$ is
naturally identified with~$C\rho_{1,0}$:
$$
\widehat{\Jet^{2}\rho}=C\rho_{1,0}.
$$

\subsection*{Acknowledgments}

The authors acknowledge partial financial support from
project BFM2002--03493.
R.\,M. wishes to thank the DURSI of the Catalan Government
for an FI grant.


\end{document}